\documentstyle[12pt,epsf]{article}
\begin{document}

\begin{flushright}
Fermilab-Pub-95/301-E \\
CDF/PUB/ELECTROWEAK/PUBLIC/3312 \\
\end{flushright}
\begin{center}
{\Large {\bf Measurement of $\sigma \cdot B(W \rightarrow e \nu)$ and
$\sigma \cdot B(Z^0 \rightarrow e^+e^-)$ in $p{\overline p}$
collisions at $\sqrt{s}=1.8$ TeV}}
\end{center}

\font\eightit=cmti8
\hfilneg
\begin{sloppypar}
\noindent
F.~Abe,$^{13}$ M.~G.~Albrow,$^7$ S.~R.~Amendolia,$^{23}$
D.~Amidei,$^{16}$
J.~Antos,$^{28}$ C.~Anway-Wiese,$^4$ G.~Apollinari,$^{26}$
H.~Areti,$^7$
M.~Atac,$^7$ P.~Auchincloss,$^{25}$
% ...
\end{sloppypar}
\vskip .025in
\begin{center}
(CDF Collaboration)
\end{center}

\vskip .025in
\begin{center}
% ...
$^4$  {\eightit University of California at Los Angeles, Los
Angeles, California  90024} \\
% ...
$^7$  {\eightit Fermi National Accelerator Laboratory, Batavia,
Illinois  60510} \\
% ...
$^{13}$ {\eightit National Laboratory for High Energy Physics (KEK),
Tsukuba,  Ibaraki 305, Japan} \\
% ...
$^{16}$ {\eightit University of Michigan, Ann Arbor, Michigan 48109}
\\
% ...
$^{23}$ {\eightit Istituto Nazionale di Fisica Nucleare, University
and Scuola Normale Superiore of Pisa, I-56100 Pisa, Italy} \\
% ...
$^{25}$ {\eightit University of Rochester, Rochester, New York 14627}
\\
$^{26}$ {\eightit Rockefeller University, New York, New York 10021}
\\
% ...
$^{28}$ {\eightit Academia Sinica, Taiwan 11529, Republic of China}
\\
% ...
\vskip 0.1in
{\bf Abstract} \\
\parbox{5in}
{We present a measurement of $\sigma \cdot B(W \rightarrow e \nu)$
and $\sigma \cdot B(Z^0 \rightarrow e^+e^-)$ in proton -
antiproton collisions at $\sqrt{s} =1.8$ TeV using a significantly
improved understanding of the integrated luminosity.  The data
represent an integrated luminosity of 19.7 pb$^{-1}$ from the
1992-1993 run with the Collider Detector at Fermilab (CDF).  We find
$\sigma \cdot B(W \rightarrow e \nu) = 2.49 \pm 0.12$~nb and $\sigma
\cdot B(Z^0 \rightarrow e^+e^-) = 0.231 \pm 0.012$~nb.  }
\end{center}

\par Measurements of the product of the production cross section and
the leptonic branching ratio for $W$ and $Z^0$ bosons, $\sigma
\cdot B(W \rightarrow e \nu)$ and $\sigma \cdot B(Z^0 \rightarrow
e^+e^-)$, test the consistency of the standard model
couplings~\cite{a_SM_ref}, the understanding of higher order QCD
contributions~\cite{nnlo_w_sigma}, and the parton distribution
functions of the proton.  In perturbation theory, the production
cross section is  predicted to  next-to-next-to-leading order (NNLO),
with $\approx$ 20\% corrections to the leading order prediction from
NLO and additional $\approx$ 3\% corrections at
NNLO~\cite{nnlo_w_sigma}.

\par In previous measurements at $\sqrt{s} = 1.8$
TeV~\cite{e_89,tau_89,mu_89}, the accuracy of the comparison to
theoretical predictions has been limited by systematic uncertainties
in the overall normalization and statistical uncertainties in the
event samples.   In Reference~\cite{r94}, CDF has presented the
measured value of the  ratio $\sigma \cdot B(W \rightarrow e
\nu)$/$\sigma \cdot B(Z^0 \rightarrow e^+e^-)$ from the 1992-1993
data, but not the individual cross sections.  In
References~\cite{elastic_prd,diff_prd,total_prd}, CDF has presented
detailed descriptions of the measurements of the elastic, single
diffractive, and total cross sections at $\sqrt{s} = 1.8$ TeV, though
not a description of the luminosity normalization.  In this Letter,
we report new measurements of $\sigma \cdot B(W \rightarrow e \nu)$
and $\sigma \cdot B(Z^0 \rightarrow e^+e^-)$ using our new
precise luminosity normalization.  Details of the luminosity
measurement are included.

\par CDF~\cite{NIM_stuff} combines a solenoidal magnetic spectrometer
with  electromagnetic (EM) and hadronic (HAD) calorimeters arranged
in a projective tower geometry covering the pseudorapidity range
$\mid \eta \mid \leq 4.2$~\cite{coords}.  Proportional chambers in
the EM shower counters provide a  measurement of shower position and
profile in both the azimuthal ($\phi$) and beam ($z$) directions.
Charged particle tracking chambers are immersed in a 1.4 T magnetic
field oriented along the beam direction.   Forward scintillator
planes known as the Beam-Beam Counters, (BBC), covering
3.2~$\leq~\mid \eta \mid~\leq$~5.9 and located 5.8 m from the nominal
interaction point, serve as the primary luminosity monitor.  For the
elastic, single diffractive, and total cross section measurements,
dedicated runs during the 1988-1989 data-taking period used a
magnetic spectrometer~\cite{elastic_prd} and forward wire
chamber telescopes~\cite{total_prd} in conjunction with the BBC.

\par For the measurement of $\sigma \cdot B(W \rightarrow e \nu)$ and
$\sigma \cdot B(Z^0 \rightarrow e^+e^-)$, $W$ and $Z^0$
candidate events are selected from a common sample of high transverse
energy electrons~\cite{r94}. The selection requires a well identified
electron candidate with transverse energy greater than 20 GeV. The
electron is required to come from a primary vertex position within 60
cm of the nominal interaction point along the $z$ direction.

\par $W$ candidates are chosen from the electron sample with the
requirement that electron candidate is well-isolated in the central
calorimeter~\cite{r94} and the missing transverse energy ($\not\!{\rm
E}_{\rm T}$), defined as the magnitude of the vector sum of
transverse energy over all calorimeter towers in the range $\mid \eta
\mid \leq 3.6$, be greater than 20 GeV.  Events which are consistent
with the $Z^0$ selection (described below) are rejected.  There
are 13796 $W$ candidate events.

\par Dielectron candidates are chosen from the electron sample with
the requirement that the first candidate electron is well isolated in
the central calorimeter and that a second isolated well identified
candidate electron~\cite{r94} also be present.  From the dielectron
sample, a $Z^0$ sample is chosen with the further requirement that
the invariant mass of the two electrons be in the range $66 - 116$
GeV/$c^2$.  There are 1312 candidate events in the $Z^0$ sample.

\par We consider backgrounds in the $W$ sample from the processes
$W\rightarrow \tau \rightarrow e$; $Z^0 \rightarrow e^+e^-$,
where one electron is not identified; $Z^0 \rightarrow \tau^+
\tau^-$, where one $\tau$ decays to a electron; mismeasured QCD jet
events; and QCD heavy flavor production.  The dominant background
contribution is from QCD processes, where one jet produces an
isolated high $p_{\rm T}$ electron candidate and the second jet is
mismeasured, mimicking $\not\!{\rm E}_{\rm T}$. The second largest
process is the background from  sequential $W \rightarrow \tau
\rightarrow e$ decays. The total background to the $W$ sample is 1700
$\pm$ 161 events~\cite{r94}.

\par We consider backgrounds in the $Z^0$ sample from QCD
processes and from the process $Z^0 \rightarrow \tau^+ \tau^-$,
where both $\tau$'s decay into electrons.   The dominant background
contribution is from QCD processes.  The total background to the
$Z^0$ sample is 21 $\pm$ 9 events~\cite{r94}.  A correction of
$(+0.5 \pm 0.2)\%$ is applied to the number of $Z^0$ candidates
to account for electron pairs in the mass window from the Drell-Yan
$\gamma$ continuum and electron pairs outside the mass window from
the $Z^0$~\cite{r94}.

\par The acceptances, which combine the fiducial and kinematic
requirements, are determined from a Monte Carlo program which
generates bosons from the lowest order diagram, $q{\overline q}
\rightarrow W$ or $Z^0$.  The bosons are given a transverse
momentum ($p_{\rm T}$) according to the $W$ $p_{\rm T}$ distribution
previously measured by CDF~\cite{boson_pt}.  The electron and
neutrino energies are smeared with the calorimeter energy
resolutions.  By varying the input parameters to the model, including
the parton distribution functions, $W$ mass, detector resolutions and
energy scale, and input boson $p_{\rm T}$ distribution, we estimate
systematic uncertainties in the acceptances.  Using the
MRSD-$^{\prime}$~\cite{mrsd} parton distribution functions and the
world averages~\cite{ewk_ave} of the ElectroWeak parameters, we find
the $W$ acceptance to be 0.342 $\pm$ 0.008 and the $Z^0$
acceptance to be 0.409 $\pm$ 0.005~\cite{r94}.

\par Electron identification efficiencies (including the trigger
efficiency) are studied with a sample of $Z^0$ candidates for
which minimal cuts have been imposed on the second lepton.  The
$Z^0$ identification efficiency is  dependent on the individual
electron efficiencies and the angular distribution of the second
electron, since the requirements and efficiencies have $\eta$
dependencies.   We find the electron selection for $W$ decays to have
an  efficiency of 0.754 $\pm$ 0.011 and the electron selection for
$Z^0$ decays to have an efficiency of 0.729 $\pm$
0.016~\cite{r94}.

\par The requirement that the primary vertex be within 60 cm of the
nominal interaction position is chosen to keep the events well
contained in the fiducial coverage.  The primary vertices have an
approximately Gaussian  distribution along the beam direction, with
$\sigma \approx$ 26 cm.  To calculate the efficiency of the vertex
cut, we model the distribution as a convolution of two Gaussians (the
$p$ and ${\overline p}$ distributions) with the accelerator $\beta$
function~\cite{beta_ref}.  The data are fit over the range $\pm$ 60
cm to give a best estimate of the accelerator parameters, which are
then used in the calculation of the efficiency.  The $p$ and
${\overline p}$ distributions are used as inputs to the calculation
on a fill-by-fill basis~\cite{fill_ref}. We estimate the uncertainty
by varying the parameters of the model within their $1\sigma$
uncertainties.  Weighting the different fills by their respective
integrated exposures, we find an efficiency of the vertex cut  of
0.955 $\pm$ 0.011.

\par The $W$ and $Z^0$ cross sections are normalized to the
visible cross section, $\sigma_{BBC}$, in the Beam Beam
counters~\cite{visible}. Hits in both planes that arrive coincident
with the particle bunches crossing through the detector serve as both
a minimum-bias trigger (the BBC trigger) and the primary luminosity
monitor.  The rate (number) of coincidences in these counters,
divided by $\sigma_{BBC}$  gives the instantaneous (integrated)
luminosity.  In previous publications, CDF normalized the BBC cross
section ($\sigma_{BBC} = 46.8 \pm 3.2$ mb) to UA4~\cite{UA4} and
accelerator measurements at $\sqrt{s} = 546$ GeV, extrapolated to
$\sqrt{s} = 1.8$ TeV~\cite{e_89}.  With recent direct measurements of
the elastic and total cross sections by
CDF~\cite{elastic_prd,total_prd}, we are able to make a direct
measurement of $\sigma_{BBC}$.

\par The value of $\sigma_{BBC}$ can be expressed as:
\begin{eqnarray}
\sigma_{BBC} = \sigma_{tot} \cdot \frac{N_{BBC}^{vis}}{N_{inel} +
N_{el}}
\end{eqnarray}
\noindent where $N_{BBC}^{vis}$ are the number of BBC triggered
events, and $N_{inel}$ and $N_{el}$ are the total number of inelastic
and elastic events.

\par For computational purposes, it is convenient to separate the
number of inelastic events into two contributions that have been
independently measured, $N_{inel} = N_i + N_d$. $N_i$ is the number
of events with a two-sided coincidence in either the BBC or the
forward telescopes, and $N_d$ is the number of events with a single
${\overline p}$ detected in the magnetic spectrometer coincident with
hits in the opposite side BBC or forward telescope~\cite{total_prd}.
We write $N_{el}$ in terms of the parameters of the fit to the
elastic scattering data, $N_{el} = A/b$, where $A=dN_{el}/dt|_{t=0}$
is the number of elastic events evaluated where the four-momentum
transfer squared, $-t$, equals zero and $b$ is the logarithmic
elastic slope parameter~\cite{elastic_prd}.  With these definitions
and the luminosity independent expression for the total cross section
as in Reference~\cite{total_prd},  $\sigma_{BBC}$ reduces to:
\begin{eqnarray}
\sigma_{BBC}=\frac{16\pi(\hbar c)^{2}} {1+\rho^{2}}
\cdot  \frac{N_{BBC}^{vis}}{N_i} \cdot \frac{A}{(A/b+N_i+N_d)^2}
\cdot N_i, \label{eqn3}
\end{eqnarray}
\noindent where $\rho$ is ratio of the real to imaginary part of the
forward scattering amplitude.  The advantage of this formulation is
that most of the quantities are measured independently, simplifying
the uncertainty calculations.

\par Table~\ref{table-sigma_values} presents the values used in the
calculation of $\sigma_{BBC}$. The quantity $N_i$ is a superset of
$N_{BBC}^{vis}$, and includes a Monte Carlo acceptance correction of
$+1.2\%$.  We find 98.7\% of $N_i$ triggered events are BBC triggered
events.  Therefore, $N_{BBC}^{vis}/N_i$ = 0.987/1.012 = 0.975, where
we have included the acceptance uncertainty in the uncertainty on
$N_i$ (the statistical uncertainty on this quantity is less than
0.1\%).  We use the measured value of $\rho = 0.140 \pm 0.069$  at
$\sqrt{s} = 1.8$ TeV~\cite{e710_rho} in our calculation.  With
Table~\ref{table-sigma_values}, these values, and
Equation~\ref{eqn3}, we calculate $\sigma_{BBC} = 51.15 \pm 1.60$ mb.

\par  In the total cross section measurement~\cite{total_prd}, we
defined a good BBC event to have hits on both sides of the detector,
coincident with beam crossing, and required the vertex position
reconstructed using timing to be within 3 m of the nominal
interaction position.  For the integrated luminosity measurement used
in $\sigma \cdot B(W \rightarrow e \nu)$ and $\sigma \cdot B(Z^0
\rightarrow e^+e^-)$, we require only that the  hits be coincident
with the beam crossing.  The difference in the event definition has
been studied in an unbiased trigger sample, where the only
requirement is a beam crossing.  The two definitions agree at the
level of 0.5\%, which is included as a systematic uncertainty for the
integrated luminosity calculation.

\par  CDF has made significant changes to the detector since the
total cross section measurements, especially in the small angle
region (removal of the forward wire chambers and a different beam
pipe).  Investigations of the vertex distributions, timing
information in the BBC's, and the rates in the EM and HAD shower
counters show no measurable difference in the BBC cross section
within a statistical uncertainty of 1\%.   Therefore we have included
an additional systematic uncertainty of 1.0\% in the normalization
to account for differences in $\sigma_{BBC}$ due to uncertainties in
the detector acceptance.

\par  The accelerator running conditions during the data taking were
also significantly different from those during the total cross
section measurements.  The average instantaneous luminosity during
the data taking was $3.5 \times 10^{30}$~cm$^{-2}$~sec$^{-1}$, in
contrast to $10^{28}$~cm$^{-2}$~sec$^{-1}$ during the total cross
section measurement.  Accidental coincidences in the BBC's (from
overlapping single diffractive events and  machine losses, for
example) have been studied in detail.  We apply an average correction
of  $(-1.3 \pm 1.0)\%$ to the integrated luminosity. We have also
investigated the effects of backgrounds which give real coincidences
(e.g., beam - gas interactions) and include a 1\% uncertainty as an
upper limit on the magnitude of these backgrounds. Combining the
measurement uncertainty with the acceptance and correction
uncertainties gives a total uncertainty of 3.6\% in the integrated
luminosity, a substantial improvement over the 6.8\% reported
previously~\cite{e_89}.  The data set for the $W$ and $Z^0$
analyses has an integrated luminosity of 19.7 $\pm$ 0.7 pb$^{-1}$.

\par
\begin{figure}
\epsfysize=7in
\epsffile[0 90 594 684]{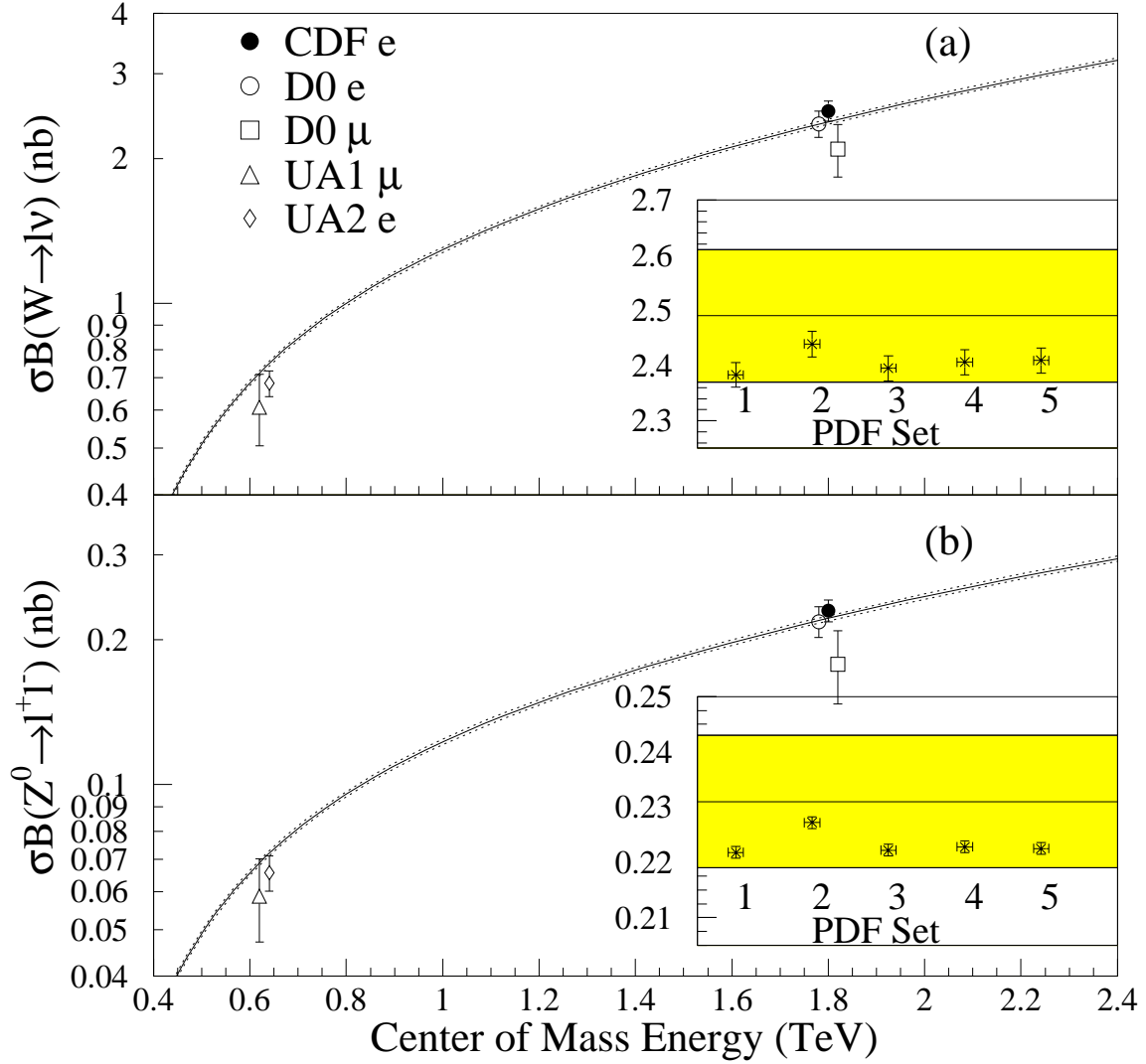}
\caption{Comparison of measured (a) $\sigma \cdot B(W \rightarrow e
\nu)$ and (b) $\sigma \cdot B(Z^0 \rightarrow e^+e^-)$ to
theoretical predictions using the calculation from Reference [2] and
MRSA [23] parton distribution functions. The UA1 and UA2 measurements
and D0 measurements are offset horizontally by $\pm$ 0.02 TeV for
clarity. In the inset,  the shaded area shows the $1\sigma$ region of
the CDF measurement; the stars show the predictions using various
parton distribution function sets (1) MRSA, (2) MRSD0$^{\prime}$, (3)
MRSD-$^{\prime}$, (4) MRSH [23] and (5) CTEQ2M [24].  The
theoretical points include a common uncertainty in the predictions
from choice of renormalization scale ($M_W/2$ to $2M_W$).}
\label{w_sigma}
\end{figure}

  Combining the event samples, backgrounds, acceptances,
efficiencies and integrated luminosities shown in
Table~\ref{e_summary}, we find  $\sigma \cdot B(W \rightarrow e\nu) =
2.49 \pm 0.02$ (stat) $\pm 0.08$ (syst) $\pm 0.09$ (lum) nb and
$\sigma \cdot B(Z^0 \rightarrow e^+ e^-) = 0.231 \pm 0.006$ (stat)
$\pm 0.007$ (syst) $\pm 0.008$ (lum) nb.  In Figure~\ref{w_sigma}, we
compare these cross section values to theoretical predictions, along
with  measurements at $\sqrt{s} = 630$ GeV from the UA1~\cite{UA1}
and UA2~\cite{UA2} collaborations  and $\sqrt{s} = 1.8$ TeV from the
D0~\cite{d0} collaboration.

\par In the insets to Figure~\ref{w_sigma}, we show the variation in
the predicted cross section times branching ratio for different sets
of parton distribution
functions~\cite{nnlo_w_sigma,stirling_94,cteq}, compared to the
current CDF measurement.   For the $W$ case, the total uncertainty is
4.9\% and is consistent with all sets, though consistently larger
than the predictions.  A recent analysis~\cite{stirling_94} of parton
distribution functions shows that the evolution of the $u$ and $d$
distributions from $Q^2 \approx$ 20 GeV$^2$ (as measured in fixed
target data) to $Q^2 \approx M_W^2$ (which determines the $W$
production cross section) depends appreciably on the gluon
distribution with $x \approx 0.05$.  In this $x$ range, the gluon
distribution is currently not well  constrained~\cite{stirling_94}
and further measurements of the $W$ and $Z$  production cross
sections could provide additional information for the parton
distribution functions.

\par In summary, we have presented measurements of $\sigma \cdot B(W
\rightarrow e \nu)$ and $\sigma \cdot B(Z^0 \rightarrow e^+e^-)$,
including a precise luminosity normalization calculation.  The
measurements are in good agreement with NNLO theoretical
predictions.  We foresee using the $\sigma \cdot B(W \rightarrow e
\nu)$ measurement as the determination of the collider luminosity in
future Tevatron runs.

\par We thank the Fermilab staff and the technical staffs of the
participating institutions for their vital contributions.  We thank
James Stirling for advice.   This work was supported by the U.S.
Department of Energy and National Science Foundation; the Italian
Istituto Nazionale di Fisica Nucleare; the Ministry of Education,
Science and Culture of Japan; the Natural Sciences and Engineering
Research Council of Canada; the National Science Council of the
Republic of China; and the A. P. Sloan Foundation.

\begin{table}[t]
\begin{center}
 \begin{tabular}{|c|c|}       \hline
&  Number of events \\ \hline \hline
$N_i$ &  208890$\pm$2558 \\
$N_d$ & 32092$\pm$1503 \\
$N_{el}$ & 78691$\pm$1463 \\ \hline
$A$ $dN_{el}/dt|_{t=0}$ &   1336532$\pm$40943 (GeV$^{-2}$) \\
$b$ Elastic Slope & 16.98$\pm$0.25 (GeV$^{-2}$) \\
Covariance$(A,b)$ & 0.93 \\ \hline
\end{tabular}
\end{center}
\vspace{-3ex}
\caption{Summary of results from the total cross section measurement
used in the calculation of $\sigma_{BBC}$ (from Reference [7,9]).  The
Covariance$(A,b)$ is the correlation coefficient in the 2
dimensional fit to the elastic slope and $dN_{el}/dt|_{t=0}$.}
\label{table-sigma_values}
\end{table}

\begin{table*}[t]
\begin{center}
\begin{tabular}{|l|c|c|}\hline
 & $W$ Events & $Z^0$ Events \\ \hline \hline
Candidates & 13796 & 1312 \\ \hline
Total Background & 1700 $\pm$ 161 & 21 $\pm$ 9 \\ \hline
Signal & 12096 $\pm$ 117 $\pm$ 161 & 1291 $\pm$ 36 $\pm$ 9 \\ \hline
Drell-Yan Correction & -- & 1.005 $\pm$ 0.002 \\ \hline
Acceptance & 0.342 $\pm$ 0.008 & 0.409 $\pm$ 0.005 \\ \hline
Efficiency & 0.754 $\pm$ 0.011 & 0.729 $\pm$ 0.016 \\ \hline
Vertex Efficiency & 0.955 $\pm$ 0.011 & 0.955 $\pm$ 0.011 \\ \hline
Luminosity & 19.7 $\pm$ 0.7 ${\rm pb}^{-1}$ & 19.7 $\pm$ 0.7 ${\rm
pb}^{-1}$
\\ \hline \hline
Cross Sections & 2.49 $\pm$ 0.02 (stat) & 0.231 $\pm$ 0.006 (stat) \\
 & $\pm$ 0.08 (syst) $\pm$ 0.09 (lum) nb
& $\pm$ 0.007 (syst) $\pm$ 0.008 (lum) nb \\
\hline
\end{tabular}
\end{center}
\caption{Summary of results on $\sigma \cdot B(W \rightarrow e \nu)$
and $\sigma \cdot B(Z^0 \rightarrow e^+e^-)$.}
\label{e_summary}
\end{table*}

\begin{thebibliography}{99}
\bibitem{coords}
The CDF coordinate system defines $z$ along the proton-beam
direction, $\theta$ as the polar angle, and $\phi$ as the azimuthal
angle.  The pseudo-rapidity is defined as $\eta =
-ln(tan(\theta/2))$.
\bibitem{fill_ref}
The Tevatron is filled with  $p$ and ${\overline p}$ beams
approximately once per day and each fill has slightly different
properties.
\bibitem{visible}
We wish to emphasize that this normalization is a visible cross
section, dependent upon the secondary interactions, photon
conversions, etc., which take place in the material between the
interaction region and the BBC's.
\end{thebibliography}
\end{document}